# Light scattering from coupled plasmonic nanospheres on a substrate

Huai-Yi Xie and Yia-Chung Chang*

*Research Center for Applied Sciences, Academia Sinica, Taipei, Taiwan 11529*

(Date: May 29, 2013)

**Abstract**

An efficient numerical method based on half-space Green's function and spherical harmonics expansion is used to study the light scattering from coupled multiple nanospheres on a substrate. The ellipsometric spectra for various geometries of coupled Au nanospheres are calculated and analyzed to realize the effects of plasmonic coupling of closely-spaced nanospheres. With only a few parameters to describe the distribution of various coupled nanosphere clusters embedded in a random distribution of nanospheres, the calculated ellipsometric spectra can fit the experimental data very well. This illustrates that our realistic model calculations can be used for determination of the distribution of nanospheres on a substrate or embedded in multilayer structures such as biological samples.

## I. Introduction

Surface plasmonic effects associated with metallic nanostructures have been wildly investigated by physicist, chemist and biologist. Surface plasmons refer to the collective charge oscillations of the free electrons at the interface between a dielectric and metal. The strong enhancement of the electric field occurs near the interface when the optical wave is strongly coupled with the free electrons near the metallic surface, and this phenomenon is called surface plasmon resonance (SPR). When we consider nanoscale metallic structures such as nanosphere, nanorod, nanoprism,…etc, the large enhancement of the electric field can arise near the surface of nanostuctures due to localized surface plasmon resonance (LSPR). The LSPR can be strongly enhanced near the junctions of aggregated plasmonic nanoparticles, such as nanoparticle dimmer [1], trimmer [2], and heptamer [3-5], and areas of strong localized electric field (so called hot spots) and the strong optical force [6] have been observed.

To observe the plasmonic coupling of aggregated nanoparticles, it is more convenient to place them on a substrate. Nanoparticle clusters can also be found in a random distribution of metallic nanoparticles on a glass substrate which was formed by dipping a chemically treated glass plate in a solution containing metallic nanoparticles of similar size, and the plasmonic effect of these nanoparticle clusters can be measured experimentally via conventional ellipsometry setup [7-8]. There were several theoretical investigations of this subject. A fully electrodynamics coupled dipole model was used to study the optical scattering from nanoparticle clusters [9]. However, this model neglects higher-order multiple contribution, which can be important when nanoparticles are near each other. The generalized Mie scattering theory which utilizes the addition theory for two-center vector spherical harmonic functions has been applied to coupled nanospheres [10-13]. An alternate method is the cluster T-matrix approach [14,15]. All the above methods are not designed to examine the effects of substrate.

Although the finite-element method (FEM) and finite domain time difference (FDTD) method [16] can be used to study the influence of the substrate on the nanoparticle clusters in general, they require very heavy computation effort to describe the sharply localized functions near plasmonic junctions, and it becomes impractical to study the combined effect of several nanoparticle clusters with various orientations. Furthermore, FDTD method cannot take into account the frequency-dependent dielectric functions accurately. Here we present an efficient method based on the half-space Green's function derived in [17] and the use spherical harmonics functions for the expansion of electromagnetic field inside the nanospheres, which is suitable for providing accurate full solution to the coupled nanospheres placed on a multilayer substrate.

In this paper, we present detailed derivations for the theory of optical scattering from clustering nanoparticles on a substrate by using the Lippmann-Schwinger (L-S) equation [7,18] to solve the electromagnetic field inside the nanospheres via expansion in terms of spherical harmonic functions. The calculated light scattering spectra for various configurations of nanoparticle clusters are analyzed to indentify the plasmonic resonances and Fano coupling effects. We further calculate the ellipsometric spectra of the combined effects of several gold nanoparticle (Au-NP) clusters with various orientations embedded in a random distribution of gold nanoparticles. By comparing our results with the corresponding experimental data with only a few fitting parameters, we can determine rather uniquely the size, average spacing, and fraction of nanoparticle clusters in a given sample. A similar comparison was reported in a previous paper [8], in which the nanoparticle clusters are resembled by pancakes with various sizes. Here, we include the realistic effects of clusters of coupled nanospheres and find more truthful description of the experimental results. In our modeling, we use the experimental dielectric constant of bulk Au. [19]

**II. Theoretical formulation of light scattering from clustering nanoparticles**

We assume that the clusters contain $N$ spheres with the same diameter on a substrate. Based on the Lippmann-Schwinger (L-S) equation for electric fields, we have

$$\mathbf{E}(\mathbf{r}) = \mathbf{E}_0(\mathbf{r}) + \sum_{i=1}^{N} (\varepsilon_i - \varepsilon_0) \int_{V_i} \mathbf{G}(\mathbf{r},\mathbf{r}') \cdot \mathbf{E}(\mathbf{r}') d^3\mathbf{r}', \qquad (1)$$

where $\mathbf{E}_0(\mathbf{r}) = e^{i\mathbf{k}_0 \cdot \mathbf{r}} \mathbf{e}_0$ is the incident electric field (a three-dimensional vector), $\varepsilon_i, i=1,2,...,N$ is the dielectric function in region $V_i, i=1,2,...,N$, $\varepsilon_0$ is the dielectric constant of the air and $\mathbf{G}(\mathbf{r},\mathbf{r}') = \frac{1}{(2\pi)^2} \int d\mathbf{k}_n e^{i\mathbf{k}_n \cdot (\boldsymbol{\rho}-\boldsymbol{\rho}')} \mathbf{g}_n(z,z')$ is the dyadic Green's function defined in Eq. (4) of Ref. [17]. Since all three components of electric field inside a uniform medium satisfy the Helmholtz equation $(\nabla^2 + k^2) E_j = 0, j=x,y,z$; thus, the solutions $E_j$ are proportional to $j_\ell(k_i r) Y_{\ell m}(\Omega)$ in the spherical coordinates [8], where $k_i$ is the wavenumber in region $V_i, i=1,2,...,N$. Note that $j_\ell$ is the spherical Bessel function of order $\ell$ and $Y_{\ell m}$ is the spherical harmonics of quantum numbers $(\ell, m)$. For convenience, we can write the electric field in the form of linear combination of localized orbitals (LCAO) (for $\mathbf{r}$ restricted in the layer where the nanoparticles reside) as follows:

$$E_\gamma(\mathbf{r}) = \sum_{\ell m} \alpha^\gamma_{i\ell m} j_\ell(k_i r) Y_{\ell m}(\Omega), \mathbf{r} \in V_i, i = 1, 2, ..., N. \quad (2)$$

where $\alpha^\gamma_{i\ell m}$ are expansion coefficients yet to be determined. Next, we use the double projection method to convert the integral equation (1) to a matrix equation for solving the unknown coefficients $\alpha^\gamma_{i\ell m}, i = 1, 2, ..., N$. First we project into the basis functions:

$j_\ell(k_i r) Y_{\ell m}$ in region $V_i$; $i = 1, 2, ..., N$. Then Eq. (1) becomes

$$S_\ell(k_p) \alpha^j_{p\ell m} = F_j^{p\ell m} + \sum_{q=1}^{N} (\varepsilon_q - \varepsilon_a) \sum_{\gamma \ell' m'} \mathbf{G}^{pq,\ell m \ell' m'}_{j\gamma} \alpha^\gamma_{q\ell' m'}, p = 1, 2, ..., N, \quad (3)$$

where $S_\ell(k_i) \equiv \int_0^a dr\, r^2 \left[j_\ell(k_i r)\right]^* j_\ell(k_i r), i = 1, 2, ..., N$, $a$ is the radius of a nanoparticle, and

$$\mathbf{G}^{pq,\ell m \ell' m'}_{j\gamma} = \sum_{n\alpha\beta} e^{-i\mathbf{k}_n \cdot \mathbf{d}_{qp}} \mathbf{S}_{n,\alpha j} \bar{\mathbf{G}}^{pq,\ell m \ell' m'}_{n,\alpha\beta} \mathbf{S}_{n,\beta\gamma}, p, q = 1, 2, ..., N. \quad (4)$$

$\mathbf{d}_{qp} \equiv \mathbf{R}_q - \mathbf{R}_p$, $\mathbf{R}_p$ is the center position of the $p$-th sphere, and $\mathbf{S}$ is a rotation matrix whose explicit form is

$$S_n = \begin{pmatrix} \cos\varphi_n & \sin\varphi_n & 0 \\ -\sin\varphi_n & \cos\varphi_n & 0 \\ 0 & 0 & 1 \end{pmatrix} = \begin{pmatrix} \sin\tilde{\varphi}_n & \cos\tilde{\varphi}_n & 0 \\ -\cos\tilde{\varphi}_n & \sin\tilde{\varphi}_n & 0 \\ 0 & 0 & 1 \end{pmatrix}, \quad (5)$$

where $\tan\varphi_n = \dfrac{k_{yn}}{k_{xn}}, \tan\tilde{\varphi}_n = \dfrac{k_{xn}}{k_{yn}}$. The matrix elements of $\bar{\mathbf{G}}^{pq,\ell m \ell' m'}_{n,\alpha\beta}$ are

$$\bar{\mathbf{G}}^{pq,\ell m \ell' m'}_{n,xx} = -\frac{q_n}{2} \tilde{u}_n \left\{ \tilde{R}_n E^{p+}_{n\ell m} \left[ B^{q+}_{n\ell' m'} + \tilde{r}_n B^{q-}_{n\ell' m'} \right] + \tilde{r}_n E^{p-}_{n\ell m} \left[ \tilde{R}_n B^{q+}_{n\ell' m'} + B^{q-}_{n\ell' m'} \right] \right\} - \frac{q_n}{2} M^{pq+}_{n\ell m \ell' m'}, \quad (6)$$

$$\bar{\mathbf{G}}^{pq,\ell m \ell' m'}_{n,xz} = -\frac{ik_n}{2} \tilde{u}_n \left\{ \tilde{R}_n E^{p+}_{n\ell m} \left[ B^{p+}_{n\ell' m'} - \tilde{r}_n B^{p-}_{n\ell' m'} \right] + \tilde{r}_n E^{p-}_{n\ell m} \left[ \tilde{R}_n B^{p+}_{n\ell' m'} - B^{p-}_{n\ell' m'} \right] \right\} - \frac{ik_n}{2} M^{pq-}_{n\ell m \ell' m'}, \quad (7)$$

$$\bar{\mathbf{G}}^{pq,\ell m \ell' m'}_{n,yy} = \frac{k_0^2}{2q_n} u_n \left\{ R_n E^{p+}_{n\ell m} \left[ B^{p+}_{n\ell' m'} + \bar{r}_n B^{p-}_{n\ell' m'} \right] + \bar{r}_n E^{p-}_{n\ell m} \left[ R_n B^{p+}_{n\ell' m'} + B^{p-}_{n\ell' m'} \right] \right\} + \frac{k_0^2}{2q_n} M^{pq+}_{n\ell m \ell' m'}, \quad (8)$$

$$\bar{\mathbf{G}}_{n,zx}^{pq\ell m\ell'm'} = -\frac{ik_n}{2}\tilde{u}_n\left\{-\tilde{R}_n E_{n\ell m}^{p+}\left[B_{n\ell'm'}^{p+} + \tilde{r}_n B_{n\ell'm'}^{p-}\right] + \tilde{r}_n E_{n\ell m}^{p-}\left[\tilde{R}_n B_{n\ell'm'}^{p+} + B_{n\ell'm'}^{p-}\right]\right\} - \frac{ik_n}{2}M_{n\ell m\ell'm'}^{pq-},$$

(9)

$$\bar{\mathbf{G}}_{n,zz}^{pq\ell m\ell'm'} = \frac{k_n^2}{2q_n}\tilde{u}_n\left\{-\tilde{R}_n E_{n\ell m}^{p+}\left[B_{n\ell'm'}^{p+} - \tilde{r}_n B_{n\ell'm'}^{p-}\right] + \tilde{r}_n E_{n\ell m}^{p-}\left[\tilde{R}_n B_{n\ell'm'}^{p+} - B_{n\ell'm'}^{p-}\right]\right\} + \frac{k_n^2}{2q_n}M_{n\ell m\ell'm'}^{pq+},$$

$$-M_{n\ell m\ell'm'}^{pq0}$$

(10)

and $F_j^{p\ell m}, p = 1, 2, ..., N$ is

$$F_j^{p\ell m} = 4\pi i^\ell E_j^0 e^{i\mathbf{k}_0 \cdot \mathbf{R}_p} \int_0^a dr\, r^2 \left[j_\ell(k_p r)\right]^* j_\ell(k_0 r)\left[e^{ik_{0z}a}Y_{\ell m}^*(\Omega_k) + R_0 e^{-ik_{0z}a}Y_{\ell m}^*(\pi - \theta_k, \varphi_k)\right],$$

(11)

where $k_0$ is the wave number for medium outside the sphere (air for the present case), $k_p = \varepsilon_p k_0, p = 1, 2, ..., N$ is the wave number inside the sphere and $q_n \equiv \sqrt{k_n^2 - k_0^2}$. $R_n$ (or $\tilde{R}_n$) denotes the reflection matrix defined in [17], including a factor $e^{-4q_n a}$. $R_0$ denotes the reflective coefficients in the grating layer without the spheres. In cylindrical coordinates, the coefficients in Eqs. (6)-(10) become

$$B_{n\ell'm'}^{q\pm} = e^{-im'\tilde{\varphi}_n}\int_{-a}^{a}dz\, e^{\pm q_n z}e^{-q_n a}I_{n\ell'm'}^q(z)$$
$$E_{n\ell m}^{p\pm} = e^{im\tilde{\varphi}_n}\int_{-a}^{a}dz\, e^{\pm q_n z}e^{-q_n a}\left[I_{n\ell m}^p(z)\right]^*,$$

(12)

where

$$I_{n\ell m}^i(z) = \sqrt{2\pi}\int_0^{\sqrt{a^2-z^2}}d\rho\,\rho\, j_\ell\left(k_i\sqrt{\rho^2+z^2}\right)\tilde{P}_\ell^m\left(\frac{z}{\sqrt{\rho^2+z^2}}\right)J_m(k_n\rho), i = 1,2,...,N$$

$\tilde{P}$ is the normalized associated Legendre function, and $J_m$ is the Bessel function of order $m$. $M_{n\ell m\ell'm'}^{pq+}$ and $M_{n\ell m\ell'm'}^{pq-}$ denote the matrix elements associated with the singular terms $e^{-q_n|z-z'|}$ and $\mathrm{sgn}(z-z')e^{-q_n|z-z'|}$, respectively. We have

$$M_{n\ell m\ell'm'}^{pq\pm} = e^{-im'\tilde{\varphi}_n}e^{im\tilde{\varphi}_n}\int_{-a}^{a}dz\left[I_{n\ell m}^p(z)\right]^*\left[\int_{-a}^{z}dz'\,e^{-q_n(z-z')}I_{n\ell'm'}^q(z') \pm \int_{z}^{a}dz'\,e^{-q_n(z'-z)}I_{n\ell'm'}^q(z')\right],$$

(13)

which satisfies the symmetry relation $\left(M_{n\ell m\ell'm'}^{pq\pm}\right)^* = \pm M_{n\ell'm'\ell m}^{qp\pm}$ when $q_n$ is real. The double integrals over $z$ and $z'$ can be reduced a single integral by using the semi-separable properties of the function $e^{-q_n|z-z'|}$ as described in Ref. [17]. $M_{n\ell m\ell'm'}^{pq0}$ is for the $\delta(z-z')$ term:

$$M_{n\ell m\ell'm'}^{pq0} = e^{-im'\tilde{\varphi}_n} e^{im\tilde{\varphi}_n} \int_{-a}^{a} dz \left[ I_{n\ell m}^{p}(z) \right]^* I_{n\ell'm'}^{q}(z), \tag{14}$$

which satisfies the symmetry relation $\left(M_{n\ell m\ell'm'}^{pq0}\right)^* = M_{n\ell'm'\ell m}^{qp0}$.

However, since the term $I_{n\ell m}^{p}(z)$ in Eq. (12) involves a fast oscillating function $J_m(k_n\rho)$, its evaluation can be time consuming with the conventional numerical integral method such as the Gauss-Legendre integration, we therefore use a polynomial expansion for the slow-varying part of the integrand:

$$j_\ell\left(k_p\sqrt{\rho^2+z^2}\right)\tilde{P}_\ell^m\left(\frac{z}{\sqrt{\rho^2+z^2}}\right) = \begin{cases} \sum_{\nu=0,2,4,\ldots}^{2\ell+4} C'_{\ell m \nu}\rho^\nu, m \text{ is even} \\ \sum_{\nu=1,3,5,\ldots}^{2\ell+5} C'_{\ell m \nu}\rho^\nu, m \text{ is odd} \end{cases}, \rho \in [0,a], \tag{15}$$

where $C'_{\ell m\nu}$ is the expansion coefficient. Then the term $I_{n\ell m}^{p}(z)$ can be expressed as:

$$I_{n\ell m}^{p}(z) = \sqrt{2\pi} \begin{Bmatrix} \sum_{\nu=0,2,4,\ldots}^{2\ell+4} \\ \sum_{\nu=1,3,5,\ldots}^{2\ell+5} \end{Bmatrix} C'_{\ell m\nu} \int_0^{\sqrt{a^2-z^2}} d\rho \rho^{\nu+1} J_m(k_n\rho)$$

$$= \sqrt{2\pi} \begin{Bmatrix} \sum_{\nu=0,2,4,\ldots}^{2\ell+4} \\ \sum_{\nu=1,3,5,\ldots}^{2\ell+5} \end{Bmatrix} C'_{\ell m\nu} \left(\sqrt{a^2-z^2}\right)^{\nu+1} \int_0^{\sqrt{a^2-z^2}} d\rho \left(\frac{\rho}{\sqrt{a^2-z^2}}\right)^{\nu+1} J_m(k_n\rho), \tag{16}$$

$$\equiv \begin{Bmatrix} \sum_{\nu=0,2,4,\ldots}^{2\ell+4} \\ \sum_{\nu=1,3,5,\ldots}^{2\ell+5} \end{Bmatrix} C_{\ell m\nu}(z) W_z(\nu+1,m)$$

where $C_{\ell m\nu}(z) \equiv \sqrt{2\pi} C'_{\ell m\nu}\left(\sqrt{a^2-z^2}\right)^{\nu+1}$ and $W_z(\nu+1,m)$ have been defined in Ref.

[17] where we can use a recursion relation to evaluate it rather than calculate the integral directly.

The sum over $n$ in Eq. (4) means an integral over in-plane wave vectors, $\mathbf{k}_n = (k_n, \tilde{\varphi}_n)$. Here, we deal with the integral over $k_n$ by using the Gaussian quadrature technique and the integral over $\tilde{\varphi}_n$ is done analytically. The results are

$$\frac{1}{2\pi}\int_0^{2\pi} d\tilde{\varphi}_n e^{i(m-m')\tilde{\varphi}_n} e^{-i\mathbf{k}_n \cdot \mathbf{d}_{qp}} = e^{-i(m-m')\varphi_{qp}} J_{m-m'}(k_n d_{qp}), \tag{17}$$

$$\frac{1}{2\pi}\int_0^{2\pi} d\tilde{\varphi}_n e^{i(m-m')\tilde{\varphi}_n} e^{-i\mathbf{k}_n \cdot \mathbf{d}_{qp}} \sin\tilde{\varphi}_n$$
$$= \frac{e^{-i(m-m')\varphi_{qp}}}{2}\left\{-i\left[J_{m-m'+1}(k_n d_{qp}) - J_{m-m'-1}(k_n d_{qp})\right]\cos\varphi_{qp}\right. \tag{18}$$
$$\left. -\left[J_{m-m'+1}(k_n d_{qp}) + J_{m-m'-1}(k_n d_{qp})\right]\sin\varphi_{qp}\right\}$$

$$\frac{1}{2\pi}\int_0^{2\pi} d\tilde{\varphi}_n e^{i(m-m')\tilde{\varphi}_n} e^{-i\mathbf{k}_n \cdot \mathbf{d}_{qp}} \cos\tilde{\varphi}_n$$
$$= \frac{e^{-i(m-m')\varphi_{qp}}}{2}\left\{\left[J_{m-m'+1}(k_n d_{qp}) + J_{m-m'-1}(k_n d_{qp})\right]\cos\varphi_{qp}\right. \tag{19}$$
$$\left. -i\left[J_{m-m'+1}(k_n d_{qp}) - J_{m-m'-1}(k_n d_{qp})\right]\sin\varphi_{qp}\right\}$$

where $\varphi_{qp} = \tan^{-1}\left(\frac{\mathbf{d}_{qp} \cdot \mathbf{e}_y}{\mathbf{d}_{qp} \cdot \mathbf{e}_x}\right)$. Finally, the coefficients $\alpha_{p\ell m}^{j}$ in Eq. (3) are solved by a direct linear solver or an iterative linear solver with quasi-minimum residue (QMR) method.

It is worth commenting that in Mie theory [20,21] (for an isolated sphere) one typically expands the electrical/magnetic filed in terms of vector spherical harmonics, taking advantage of the nature for transverse waves. However, for spheres on a substrate, it is more convenient to expand all three Cartesian components of the electrical/magnetic filed in terms of scalar spherical harmonics, which makes the implementation much simpler and computationally more efficient. The equivalence between the two approaches is demonstrated analytically in Appendix A.

**III. Comparison with the generalized Mie scattering theory for isolated clusters**

To test the numerical accuracy of our method, we compare our calculated results with those obtained by the generalized Mie scattering theory [10-13] for coupled nanospheres without a substrate. Fig. 2 shows electric field at the top (the position $O$ with $z = 0$) of trimers, $|\mathbf{E}|$ as a function of photon energy for light scattering from a trimer of Au spheres of the same diameter $d = 80nm$ (Fig. 1(b) with $\varphi_d = 0$) with two different gaps: (a) gap $g = 10 nm$ and (b) gap $g = 2 nm$, obtained by the present Green's function method (dashed) and the generalized Mie scattering theory (solid). In our calculation, the cutoff of angular momentum quantum number $\ell$ ($\ell_c$), the number of $k_n$ meshes ($N_k$), and the number of $z$ meshes ($N_z$) used are : (a) ($\ell_c$, $N_k$, $N_z$) = (6,101,100) and (b) ($\ell_c$, $N_k$, $N_z$) = (9,121,100), which have been tested to ensure convergent results.

We noticed from Fig. 2 when the gap between nanoparticles is small, a stronger coupling effect between the aggregations occurs, which leads to large enhancement of local field for photon energy near the plasmonic resonance. This strong coupling effect also leads to a red shift of the LSPR resonance spectrum. As shown in this figure, our Green's function results agree well with the generalized Mie scattering theory in the absence of a substrate.

## IV. Ellipsoemtric spectra of isolated clusters for specular and off-spectacular reflections

We present the calculated reflectance, $R_s = |r_s|^2$ and $R_p = |r_p|^2$ for s- and p-polarized light, where $r$ denotes the Fresnel reflection coefficient for isolated clusters of Au-NPs (with $d = 80$nm and $\varphi_d = 0$ in Fig. 1) on glass substrate with three different angles of incidence: $55^o$ (green), $60^o$ (red) and $65^o$ (blue) for specular reflection (Fig. 3) and off-specular reflection (Fig. 4), normalized to an area $A_{\text{cell}} = p^2$. For the off-specular reflection, the scattered light is detected at an angle normal to the surface of the substrate. This is a better geometry for microscopic ellipsometry measurements due to the ease of focusing the objective. The gap distance (*g*) between nanoparticles adopted is 2nm and the normalization area is taken to be $A_{cell} = 530\text{nm} \times 530\text{nm}$ (which is roughly the pixel size considered in a microscopic imaging ellipsometry). We use ($\ell_c$, $N_k$, $N_z$) = (8,71,100) in all cases in order to ensure convergent results.

Comparing Figs. 3 and 4, we find that the value of $R_s$ in the off-specular reflection (over the entire energy range) is much smaller than in the specular reflection, which will lead to a large amplitude ratio, $\Psi$ for the off-specular reflection. This is direct consequence of the interference effect. For the specular reflection, the optical paths of light scattered from all naoparticels in the cluster are the same (i.e. phase coherent), while for off-specular reflection, there is a variation of optical path given by $d_j \sin \theta$, where *d*$_j$ is the location of the center of the *j*-th nanosphere from the origin along the x-axis and $\theta$ is the angle of incidence. Thus, for off-specular reflection, a partially destructive interference can occur when

$d_e \sin\theta = \lambda/2$ where $d_e$ is the effective separation between end nanoparticles in the cluster and $\lambda$ is the wavelength. Such an interference pattern is clearly seen in the $R_s$ spectra for off-specular reflection (see Fig. 4) with dips occurring at wavelengths approximately matching the condition, $\lambda = 2d_e \sin\theta$. The degree of destructive interference depends mainly on the arrangement of nanoparticles. For the dimer case and $\theta = 65^0$, the wavelength location of the dip in the $R_s$ spectra is approximately equal to 7.1eV. Hence $d_e$ satisfying the condition $2d_e \sin 65^0 = 7.1eV$ leads to an effective separation of 96.4nm. Similarly, we obtain the effective separation lengths around 140nm and 190 nm, respectively for a chain of 3 and 4 nanospheres. In Fig. 4, we also noticed that the destructive interference in trimer and heptamer is much weaker than in the chain of nanoparticles.

It is well known that near the plasmonic resonance the electric field normal to the metallic surface is localized and greatly enhanced. In our studies, the substrate lies in the *x-y* plane and the plane of incidence is in the *x-z* plane. For the s-polarized light, the electric field of the incident light is along the *y* axis. Thus, near plasmonic resonance $E_y$ is enhanced near the side edges of nanoparticles and localized near *x* = $d_j$. For the p-polarized light, the magnetic field of the incident light is along the *y* axis. Thus $H_y$ near the entire circumference of nanoparticles in the *x-z* plane is enhanced. Thus, for chains of nanoparticles we find strong plasmoinc coupling between neighboring nanoparticles for the p-polarized light but not for the s-polarized light, which leads a splitting of the plasmonic peak as seen in Figs. 3 and 4. On the other hand, for the heptamer, strong plasmoinc coupling between neighboring nanoparticles exists for both the s- and p-polarized light (since the heptamer contains chains along both *x* and *y* directions) as can be seen in Figs. 3(e) and 4(e), where a splitting is of the plasmonic peak is seen for both $R_s$ and $R_p$. These features would allow easy identification of the arrangements of metallic nanoparticle clusters via microscopic SE even though the image of the cluster itself cannot be resolved due to their sub-wavelength scale.

**V. Ellipsometric spectra of randomly oriented isolated clusters**

Fig. 5 show the calculated ellipsometric spectra ($\Psi$ and $\Delta$) for isolated clusters of Au-NPs (with $d = 80$nm) on glass substrate with random orientation. Three different angles of incidence: $55^o$ (solid lines), $60^o$ (dashed lines) and $65^o$ (dash-dotted lines) are considered. The gap distance (*g*) between nanoparticles adopted is 2nm and the normalization area used for these Au-NP clusters used is $A_{cell} = 245\text{nm} \times 245\text{nm}$. In order to ensure convergent results, we use ($\ell_c$, $N_k$, $N_z$) = (8,71,100) in all cases. To model the effect of random orientation, we average over the azimuth angle $\varphi_d$ as shown in Fig. 1 by using 10 sampling values for $\varphi_d$ evenly divided within the irreducible segment (determined by point group symmetry). As expected, we find that the plasmonic resonance peak splits into multiple peaks accompanying with some dip structure due to the strong coupling and Fano effect between adjacent nanospheres. The degree of enhancement due to the coupling effect depends on the arrangement of nanoparticles, the adjacent distance between nanopaticles, and the particle size.

Comparing Fig. 5(a) and Fig. 5(d), we noticed that the ψ spectra of the two cases are quite similar. This can be attributed to the fact that after average over azimuthal angle $\varphi_d$, both cases have the same effective diameter (~ $2d+g$), where $g$ denotes the gap distance between adjacent nanospheres. However, the Δ spectra of the two cases are quite different, indicating that the Δ is sensitive to the arrangement of nanoparticles while ψ is not. Similarly, Figs. 5(b) and 6(e) have very similar ψ spectra but different Δ spectra, again because the two cases have the same effective diameter (~ $3d+2g$), giving further confirmation to the above argument.

## VI. Theoretical formulation of light scattering from clusters embedded in a random distribution of non-aggregated nanoparticles

In our pervious work [8], we modeled clustering nanoparticles as pancakes with different diameters (the equivalent spheroid model), which are embedded in a random distribution of non-aggregated nanoparticles. Here, we treat the clusters as realistic aggregations of nanoparticles which includes the strong plasmonic coupling effect between the closely-spaced nanoparticles. In order to describe the clusters embedded in a random distribution of non-aggregated nanoparticles, we use three types of local functions similar to our pervious work [8]: $\mathbf{u}_1(\mathbf{r})$ for nonclustering nanoparticles which are randomly distributed nanoparticles with identical diameters, $\mathbf{c}_\alpha(\mathbf{r})$ for small realistic aggregation with several different types labeled by $\alpha$ and $\mathbf{p}(\mathbf{r})$ for large patch of clustering nanoparticles. $N_u, N_c$ and $N_p$ denotes the numbers of cells occupied by non-aggregated nanoparticles, clustering nanoparticles and large patch of clustering nanoparticles, respectively. The fractions of areas occupied by non-aggregated nanoparticles, clustering nanoparticles and large patch of clustering nanoparticles are denoted by $f_c = N_c/N$  $f_p = N_p/N$  and  $f_u = N_u/N = 1 - f_c - f_p$, respectively. The L-S equation for the local function $\mathbf{u}_1(\mathbf{r})$ is given by [8]

$$\mathbf{u}_1(\mathbf{r}) = \sqrt{N}\mathbf{E}_0(\mathbf{r}) + \int\frac{d\phi_n}{(2\pi)^2}\int k_n dk_n S(\mathbf{K}_n)\int d\mathbf{r}' e^{i\mathbf{k}_n\cdot(\boldsymbol{\rho}-\boldsymbol{\rho}')}\mathbf{g}_n(z,z')\cdot V_1(\mathbf{r}')\mathbf{u}_1(\mathbf{r}'), \quad (20)$$

where $\mathbf{K}_n = \mathbf{k}_n - \mathbf{k}_0$ and $V_1(\mathbf{r}') = (\varepsilon_a - 1)(\omega/c)^2$ for $\mathbf{r}'$ inside the nanoparticle and vanishes otherwise and $S(\mathbf{K}_n)$ is the structure factor defined as

$$S(\mathbf{K}_n) = \frac{1}{N}\sum_i\sum_j e^{-i\mathbf{K}_n\cdot(\mathbf{R}_j-\mathbf{R}_i)} = 1 + fS_1(\mathbf{K}_n), \quad (21)$$

$$S_1(\mathbf{K}_n) = \sum_{j\neq 1} e^{-i\mathbf{K}_n\cdot(\mathbf{R}_j-\mathbf{R}_1)} e^{-(\mathbf{R}_j-\mathbf{R}_1)^2/2\lambda_c^2} \approx \frac{2\pi\lambda_c^2}{A_{cell}} e^{-K_n^2\lambda_c^2/2} - \frac{2\pi}{A_{cell}}\int_0^{R_u} J_0(K_n R) e^{-R^2/2\lambda_c^2} R\,dR,$$

(22)

where $f$ is the similarity factor describing the average ratio of local wave functions associated with nanoparticles on different sites in a given distribution, $R_u$ is the cell radius which relates to the cell area by $A_{cell} = \pi R_u^2$, and $\lambda_c$ is the coherent length of light [20]. The local function $\mathbf{p}(\mathbf{r})$ for a large patch of nanoparticles also satisfy Eq. (20), but with the cell area replaced by $A_{cell} = \sqrt{3}d^2/2$. The local functions for clusters $\mathbf{c}_\alpha(\mathbf{r})$ are described by Eqs. (2) and (3) and they are solved according to Eq. (1). For simplicity and to reduce computation time, we regard clustering nanoparticles as the combination of chains of nanoparticles (Fig. 1(a)) including $\beta$ nanospheres ($\beta = 2,3,4$) and close-packed clustering nanoparticles such as trimer (Fig. 1(b)) and heptamer (Fig. 1(c)) in the x-y plane with various orientations. The clusters are classified by different types, labeled by $\alpha$ according to their corresponding effective diameter, $d_\alpha$ after averaging over orientation. The weighting factors used for chains of nanoparticles with effective diameters $2d+g$, $3d+2g$, and $4d+3g$ (where g is gap distance) are $\gamma p_2$, $\gamma p_3$, and $p_4$, respectively, and those for trimer (with effective diameters $2d$) and heptamer (with effective diameters $3d$) are $(1-\gamma)p_2$ and $(1-\gamma)p_3$, where $\gamma$ is the fraction (between 0 and 1) describing the contribution of chains of clusters for a given cluster size. $p_\alpha = \frac{1}{3}f_c/\eta_\alpha$ denotes the weighting factor for clusters of type $\alpha$ such that $\sum_{\alpha=1}^{3} p_\alpha\eta_\alpha = f_c$, where $\eta_\alpha = A_\alpha/A_{cell}$ with $A_\alpha$ denotes the circular area occupied by a cluster of type $\alpha$. For simplicity, other possible cluster arrangements with effective diameter close to $4d$ have been neglected. The gap distance ($g$) between nanoparticles should be close enough to realize the coupling effect between them. Furthermore, we average over the azimuth angle $\varphi_d$ as indicated in Fig. 1. Note that in our pervious work [8], the equivalent spheroid model does not obtain any information of the coupling effect of the closely-space nanoparticles. Instead, the coupling effect between several types of equivalent spheroids with different diameters with the surrounding non-aggregated nanoparticles was included to produce the desired effect [8]. Note that Eqs. (5) and Eq. (14) in Ref. [8] are also solved within the spherical harmonic basis similar to Eq. (3). The principal-order reflection coefficient is obtained by Eq. (15) in Ref. [8].

## VII. Ellipsometric spectra of randomly distributed nanoparticles with clusters

In a previous work [8], we have presented the calculated ellipsometric spectra ($\Psi$ and $\Delta$) by using the spherical harmonics-based Green's function (SHGF) method without considering the clustering effect of nanoparticles. (See Fig. 4 of [8]) The fitting parameters used to produce a good fit include the particle diameter ($d$), particle height ($h$), average inter-particle distance ($p$), the similarity factor ($f$), and the coherent length of light ($\lambda_c$). Here, we found that similar quality of fit can be obtained by using $h = d$ (i.e. spherical instead of ellipsoidal nanoparticles); thus, one less fitting parameter is used. The coherent length used is 3500nm [22], which is reasonable for the light sources adopted in the J. A. Wollam VUV ellipsometer. The results are not very sensitive to the coherence length as long as it is between 3000nm and 5000nm. Furthermore, the particle diameters used to get the best fit are kept the same as the nominal value as specified, except for the $d$=40nm case (for which we need to use a 10% smaller size to obtain a good fit). The new fitting results are shown in Fig. 6 with the best-fit parameters given in Table 1. These results are similar to Fig. 4 of our previous work [8], but here we imposed the constraint $h = d$ and enlarged the set of basis functions (with $\ell_c$=6) to ensure convergence.

Next, we describe the clustering effect on the SE measurements. Instead of using the equivalent pancake-like spheroid model as in Ref. [8], here we use the more realistic aggregated-nanosphere model to include the coupling effect between the closely-spaced nanoparticles. For clusters of effective diameters around $2d$ and $3d$, we assume a fraction ($\gamma = 0.7$) of the clusters is chain-like and the remainder is close-packed. For clusters of effective diameters around $4d$, only chain-like clusters are considered. All other larger clusters are modeled by a large patch of close-packed nanoparticle with a weighting factor $f_\mathrm{p}$. The gap distance between adjacent nanoparticles are taken to 0.025*$d$, which is small enough to simulate the coupling effect. (Any smaller value will not affect the fitting results, but requires longer computation time to converge) Here we have neglected the coupling between clusters and the surrounding (random distribution of identical nanoparticles) since the major enhancement has already been included in the coupling between the closely-spaced nanoparticles. The reflectivities of these small clusters averaged over all orientations are then mixed with the results for surrounding random distribution of nanoparticles and large patch to calculate $\Psi$ and $\Delta$. The results are shown in Fig. 7, along with experimental data (taken from Ref. [8]) for comparison. The best-fit values for weighting factors $f_c$ and $f_p$ and the mean-square-errors (MSE) for both $\psi$ and $\Delta$ (in parentheses) are also listed in Table 1. The multiple-peak feature accompanying a dip structure agrees reasonably well with experiment. It is noted that the fit is much improved (with significantly smaller MSE) when the clustering effect is included for photon energies near the plasmonic resonances. The quality of fit is comparable to the one shown in Fig. 6 of Ref. [8], but here we have kept the aspect ratio *h/d*=1 for nanoparticles and the clustering model used is a more truthful representation of the actual system.

## VIII. Conclusion

In conclusion, we have used a spherical harmonics-based Green's function (SHGF) method for studying the plasmonic coupling effects in a cluster of metallic nanoparticles. We show the SE spectra have features that can be related to the arrangements of nanoparticles in a given cluster. This point to an application that one can use the microscopic ellipsometry to determine the precise arrangements of metallic nanoparticles in a cluster with nanoscale resolution even though the image itself may not be resolved clearly due to the small size of nanoparticle. The method is also applied to the metrology of clustering nanoparticles embedded in a random distribution of nanoparticles on multilayer films. Rather than using the equivalent spheroid model to describe clustering nanoparticles as in Ref. [8], here we model the clusters as the realistic aggregation of nanospheres which include the plasmonic coupling effect between closely-spaced nanoparticles. Base on this realistic model, we can uniquely provide structure information of nanoparticles on the substrate, including the average particle size (*d*), average pitch (*p*), fraction of small clusters ($f_c$), and fraction of nanoparticles in large patches ($f_p$) within a selected spot (e.g. a 10μm✕10μm area) of a given image. We believe our efficient and realistic modeling is very useful for nondestructive metrology of nanoparticles covered samples.

**Acknowledgements**

The authors thank Shih-Hsin Hsu and Guangwei Li for fruitful discussions. This work was supported in part by the Nanoproject of Academia Sinica and National Science Council of the Republic of China under Contract No. NSC

## Appendix A

In this appendix, we derive the relations between the vector spherical harmonics used in Mie Theory [20] and the scalar harmonic functions in Cartesian frame. In the Mie Theory [20], the two vector spherical harmonic function $\mathbf{M}_{o\ell m}$ and $\mathbf{N}_{e\ell m}$ are expressed by [21]

$$\mathbf{M}_{o\ell m} = \frac{m}{\sin\theta}\cos m\phi P_\ell^m(\cos\theta) j_\ell(\rho)\mathbf{e}_\theta - \sin m\phi \frac{dP_\ell^m(\cos\theta)}{d\theta} j_\ell(\rho)\mathbf{e}_\phi, \qquad (A\text{-}1)$$

$$\mathbf{N}_{e\ell m} = \ell(\ell+1)\cos m\phi P_\ell^m(\cos\theta)\frac{j_\ell(\rho)}{\rho}\mathbf{e}_r + \cos m\phi \frac{dP_\ell^m(\cos\theta)}{d\theta}\frac{1}{\rho}\frac{d[\rho j_\ell(\rho)]}{d\rho}\mathbf{e}_\theta$$
$$-m\sin m\phi \frac{P_\ell^m(\cos\theta)}{\sin\theta}\frac{1}{\rho}\frac{d[\rho j_\ell(\rho)]}{d\rho}\mathbf{e}_\phi$$

$$(A\text{-}2)$$

Writing the unit vector in the spherical coordinate into their Cartesian components, we obtain

$$\begin{pmatrix}\mathbf{e}_r\\ \mathbf{e}_\theta\\ \mathbf{e}_\phi\end{pmatrix} = \begin{pmatrix}\sin\theta\cos\phi & \sin\theta\sin\phi & \cos\theta\\ \cos\theta\cos\phi & \cos\theta\sin\phi & -\sin\theta\\ -\sin\phi & \cos\phi & 0\end{pmatrix}\begin{pmatrix}\mathbf{e}_x\\ \mathbf{e}_y\\ \mathbf{e}_z\end{pmatrix}. \qquad (A\text{-}3)$$

First, we discuss the x-component of the function $\mathbf{M}_{o\ell m}$. The recursion relationships of associate Legendre functions are

$$\frac{m}{\sin\theta}P_\ell^m(\cos\theta)\cos\theta = -\frac{1}{2}\left[P_\ell^{m+1}(\cos\theta)+(\ell-m+1)(\ell+m)P_\ell^{m-1}(\cos\theta)\right], \qquad (A\text{-}4)$$

$$\frac{dP_\ell^m(\cos\theta)}{d\theta} = -\frac{1}{2}\left[(\ell-m+1)(\ell+m)P_\ell^{m-1}(\cos\theta)-P_\ell^{m+1}(\cos\theta)\right]. \qquad (A\text{-}5)$$

Using the product-to-sum identities:

$$\cos m\phi\cos\phi = \frac{1}{2}\left[\cos(m+1)\phi+\cos(m-1)\phi\right], \qquad (A\text{-}6)$$

and

$$\sin m\phi\sin\phi = \frac{1}{2}\left[\cos(m-1)\phi-\cos(m+1)\phi\right], \qquad (A\text{-}7)$$

We obtain for the x-component of the function, $M_{o\ell m}^x$:

$$M_{o\ell m}^x = -\frac{j_\ell(\rho)}{4}\Big[C_\ell^{m+1}Y_\ell^{m+1}+(-1)^{m+1}C_\ell^{m+1}Y_\ell^{-(m+1)}$$
$$+(\ell-m+1)(\ell+m)C_\ell^{m-1}Y_\ell^{m-1}+(\ell-m+1)(\ell+m)(-1)^{m-1}C_\ell^{m-1}Y_\ell^{-(m-1)}\Big], \qquad (A\text{-}8)$$

where the coefficient $C_\ell^m = \sqrt{\dfrac{4\pi}{2\ell+1}\dfrac{(\ell+m)!}{(\ell-m)!}}$.

Second, we discuss the y-component of the function $\mathbf{M}_{o\ell m}$. Using the product-to-sum identities

$$\cos m\phi \sin\phi = \frac{1}{2}\left[\sin(m+1)\phi - \sin(m-1)\phi\right], \tag{A-9}$$

$$\sin m\phi \cos\phi = \frac{1}{2}\left[\sin(m+1)\phi + \sin(m-1)\phi\right], \tag{A-10}$$

and Eq. (A-4) and Eq. (A-5), we obtain for y-component of the function, $M_{o\ell m}^y$:

$$\begin{aligned}M_{o\ell m}^y = \frac{j_\ell(\rho)}{4i}&\left[-C_\ell^{m+1}Y_\ell^{m+1} + (-1)^{m+1} C_\ell^{m+1}Y_\ell^{-(m+1)}\right.\\&\left.+(\ell-m+1)(\ell+m)C_\ell^{m-1}Y_\ell^{m-1} - (\ell-m+1)(\ell+m)(-1)^{m-1} C_\ell^{m-1}Y_\ell^{-(m-1)}\right]\end{aligned}. \tag{A-11}$$

Similarly, for the z-component of the function $\mathbf{M}_{o\ell m}$ we obtain

$$M_{o\ell m}^z = -\frac{m}{2}C_\ell^m\left[Y_\ell^m + (-1)^m Y_\ell^{-m}\right]j_\ell(\rho). \tag{A-12}$$

In the case of $\mathbf{N}_{e\ell m}$, the three Cartesian components are

$$\begin{aligned}N_{e\ell m}^x &= \frac{\ell(\ell+1)}{2\ell+1}\cos m\phi \cos\phi \sin\theta P_\ell^m(\cos\theta)\left[j_{\ell-1}(\rho) + j_{\ell+1}(\rho)\right]\\&+ \frac{1}{2\ell+1}\cos m\phi \cos\phi \cos\theta \frac{dP_\ell^m(\cos\theta)}{d\theta}\left[(\ell+1)j_{\ell-1}(\rho) - \ell j_{\ell+1}(\rho)\right]\\&+ \frac{1}{2\ell+1}\sin m\phi \sin\phi \frac{P_\ell^m(\cos\theta)}{\sin\theta}\left[(\ell+1)j_{\ell-1}(\rho) - \ell j_{\ell+1}(\rho)\right]\\&\equiv j_{\ell-1}(\rho)N_{e\ell m}^{x(1)}(\Omega) + j_{\ell+1}(\rho)N_{e\ell m}^{x(2)}(\Omega)\end{aligned} \tag{A-13}$$

$$\begin{aligned}N_{e\ell m}^y &= \frac{\ell(\ell+1)}{2\ell+1}\cos m\phi \sin\phi \sin\theta P_\ell^m(\cos\theta)\left[j_{\ell-1}(\rho) + j_{\ell+1}(\rho)\right]\\&+ \frac{1}{2\ell+1}\cos m\phi \sin\phi \cos\theta \frac{dP_\ell^m(\cos\theta)}{d\theta}\left[(\ell+1)j_{\ell-1}(\rho) - \ell j_{\ell+1}(\rho)\right]\\&- \frac{1}{2\ell+1}\sin m\phi \cos\phi \frac{P_\ell^m(\cos\theta)}{\sin\theta}\left[(\ell+1)j_{\ell-1}(\rho) - \ell j_{\ell+1}(\rho)\right]\\&\equiv j_{\ell-1}(\rho)N_{e\ell m}^{y(1)}(\Omega) + j_{\ell+1}(\rho)N_{e\ell m}^{y(2)}(\Omega)\end{aligned} \tag{A-14}$$

$$\begin{aligned}N_{e\ell m}^z &= \frac{\ell(\ell+1)}{2\ell+1}\cos m\phi \cos\theta P_\ell^m(\cos\theta)\left[j_{\ell-1}(\rho) + j_{\ell+1}(\rho)\right]\\&- \frac{1}{2\ell+1}\cos m\phi \sin\theta \frac{dP_\ell^m(\cos\theta)}{d\theta}\left[(\ell+1)j_{\ell-1}(\rho) - \ell j_{\ell+1}(\rho)\right],\\&\equiv j_{\ell-1}(\rho)N_{e\ell m}^{z(1)}(\Omega) + j_{\ell+1}(\rho)N_{e\ell m}^{z(2)}(\Omega)\end{aligned} \tag{A-15}$$

where we have used the recursion relations for the spherical Bessel functions

$$j_\ell(\rho) = \frac{\rho}{2\ell+1}\left[j_{\ell-1}(\rho) + j_{\ell+1}(\rho)\right], \tag{A-16}$$

$$\frac{dj_\ell(\rho)}{d\rho} = \frac{1}{2\ell+1}\left[\ell j_{\ell-1}(\rho) - (\ell+1) j_{\ell+1}(\rho)\right]. \tag{A-17}$$

Because there is no suitable recursion for Eqs. (A-13)-(A-15), we write

$$N_{e\ell m}^{i(p)} = \sum_{\ell' m'} c_{\ell m;\ell' m'}^{i(p)} Y_{\ell'}^{m'}, i = x, y, z;\ p = 1, 2. \tag{A-18}$$

Multiply the conjugate of the spherical harmonic function on both side in the above equation and integrate over the angules $\theta$ and $\phi$, we have

$$\int d\Omega N_{e\ell m}^{i(p)} \left[Y_{\ell''}^{m''}\right]^* = \int d\Omega \sum_{\ell' m'} c_{\ell m;\ell' m'}^{i(p)} Y_{\ell'}^{m'} \left[Y_{\ell''}^{m''}\right]^* = c_{\ell m;\ell'' m''}^{i(p)}, i = x, y, z;\ p = 1, 2. \tag{A-19}$$

Thus, we can rewrite Eqs. (A-13)-(A-15) straightforwardly as follows:

$$N_{e\ell m}^x = j_{\ell-1}(\rho) N_{e\ell m}^{x(1)}(\Omega) + j_{\ell+1}(\rho) N_{e\ell m}^{x(2)}(\Omega)$$

$$= \begin{cases} j_{\ell-1}(\rho) \sum_{\substack{m'=\pm 1,\pm 3,\ldots \\ |m'|<\ell-1}} c_{\ell m;\ell-1,m'}^{x(1)} Y_{\ell-1}^{m'} + j_{\ell+1}(\rho) \sum_{\substack{m'=\pm 1,\pm 3,\ldots \\ |m'|<\ell+1}} c_{\ell m;\ell+1,m'}^{x(2)} Y_{\ell+1}^{m'}, & \text{if } m \text{ is even} \\ j_{\ell-1}(\rho) \sum_{\substack{m'=\pm 2,\pm 4,\ldots \\ |m'|<\ell-1}} c_{\ell m;\ell-1,m'}^{x(1)} Y_{\ell-1}^{m'} + j_{\ell+1}(\rho) \sum_{\substack{m'=\pm 2,\pm 4,\ldots \\ |m'|<\ell+1}} c_{\ell m;\ell+1,m'}^{x(2)} Y_{\ell+1}^{m'}, & \text{if } m \text{ is odd} \end{cases}, \tag{A-20}$$

$$N_{e\ell m}^y = j_{\ell-1}(\rho) N_{e\ell m}^{y(1)}(\Omega) + j_{\ell+1}(\rho) N_{e\ell m}^{y(2)}(\Omega)$$

$$= \begin{cases} j_{\ell-1}(\rho) \sum_{\substack{m'=\pm 1,\pm 3,\ldots \\ |m'|<\ell-1}} c_{\ell m;\ell-1,m'}^{y(1)} Y_{\ell-1}^{m'} + j_{\ell+1}(\rho) \sum_{\substack{m'=\pm 1,\pm 3,\ldots \\ |m'|<\ell+1}} c_{\ell m;\ell+1,m'}^{y(2)} Y_{\ell+1}^{m'}, & \text{if } m \text{ is even} \\ j_{\ell-1}(\rho) \sum_{\substack{m'=\pm 2,\pm 4,\ldots \\ |m'|<\ell-1}} c_{\ell m;\ell-1,m'}^{y(1)} Y_{\ell-1}^{m'} + j_{\ell+1}(\rho) \sum_{\substack{m'=\pm 2,\pm 4,\ldots \\ |m'|<\ell+1}} c_{\ell m;\ell+1,m'}^{y(2)} Y_{\ell+1}^{m'}, & \text{if } m \text{ is odd} \end{cases}, \tag{A-21}$$

$$N_{e\ell m}^z = j_{\ell-1}(\rho) N_{e\ell m}^{z(1)}(\Omega) + j_{\ell+1}(\rho) N_{e\ell m}^{z(2)}(\Omega)$$

$$= \begin{cases} j_{\ell-1}(\rho) \sum_{\substack{m'=\pm 1,\pm 3,\ldots \\ |m'|<\ell-1}} c_{\ell m;\ell-1,m'}^{z(1)} Y_{\ell-1}^{m'} + j_{\ell+1}(\rho) \sum_{\substack{m'=\pm 1,\pm 3,\ldots \\ |m'|<\ell+1}} c_{\ell m;\ell+1,m'}^{z(2)} Y_{\ell+1}^{m'}, & \text{if } m \text{ is odd} \\ j_{\ell-1}(\rho) \sum_{\substack{m'=\pm 2,\pm 4,\ldots \\ |m'|<\ell-1}} c_{\ell m;\ell-1,m'}^{z(1)} Y_{\ell-1}^{m'} + j_{\ell+1}(\rho) \sum_{\substack{m'=\pm 2,\pm 4,\ldots \\ |m'|<\ell+1}} c_{\ell m;\ell+1,m'}^{z(2)} Y_{\ell+1}^{m'}, & \text{if } m \text{ is even} \end{cases}. \tag{A-22}$$

Furthermore, we have

$$\mathbf{M}_{o\ell m} = -\frac{j_\ell(\rho)}{4}\Big[C_\ell^{m+1}Y_\ell^{m+1} + (-1)^{m+1}C_\ell^{m+1}Y_\ell^{-(m+1)}$$
$$+(\ell-m+1)(\ell+m)C_\ell^{m-1}Y_\ell^{m-1} + (\ell-m+1)(\ell+m)(-1)^{m-1}C_\ell^{m-1}Y_\ell^{-(m-1)}\Big]\mathbf{e}_x$$
$$+\frac{j_\ell(\rho)}{4i}\Big[-C_\ell^{m+1}Y_\ell^{m+1} + (-1)^{m+1}C_\ell^{m+1}Y_\ell^{-(m+1)}$$
$$+(\ell-m+1)(\ell+m)C_\ell^{m-1}Y_\ell^{m-1} - (\ell-m+1)(\ell+m)(-1)^{m-1}C_\ell^{m-1}Y_\ell^{-(m-1)}\Big]\mathbf{e}_y, \quad \text{(A-23)}$$
$$-\frac{m}{2}C_\ell^m\Big[Y_\ell^m + (-1)^m Y_\ell^{-m}\Big]j_\ell(\rho)\mathbf{e}_z$$
$$\equiv \boldsymbol{\alpha}_1 j_\ell(\rho)Y_\ell^{m+1} + \boldsymbol{\alpha}_2 j_\ell(\rho)Y_\ell^{-(m+1)} + \boldsymbol{\alpha}_3 j_\ell(\rho)Y_\ell^{m-1} + \boldsymbol{\alpha}_4 j_\ell(\rho)Y_\ell^{-(m-1)}$$
$$+\boldsymbol{\alpha}_5 j_\ell(\rho)Y_\ell^m + \boldsymbol{\alpha}_6 j_\ell(\rho)Y_\ell^{-m}$$

$$\mathbf{N}_{e\ell m} = \Big[j_{\ell-1}(\rho)N_{e\ell m}^{x(1)}(\Omega) + j_{\ell+1}(\rho)N_{e\ell m}^{x(2)}(\Omega)\Big]\mathbf{e}_x$$
$$+\Big[j_{\ell-1}(\rho)N_{e\ell m}^{y(1)}(\Omega) + j_{\ell+1}(\rho)N_{e\ell m}^{y(2)}(\Omega)\Big]\mathbf{e}_y + \Big[j_{\ell-1}(\rho)N_{e\ell m}^{z(1)}(\Omega) + j_{\ell+1}(\rho)N_{e\ell m}^{z(2)}(\Omega)\Big]\mathbf{e}_z$$

$$= \begin{cases} j_{\ell-1}(\rho)\Bigg[\mathbf{e}_x \sum_{\substack{m'=\pm 1,\pm 3,\ldots \\ |m'|<\ell-1}} c_{\ell m;\ell-1,m'}^{x(1)} Y_{\ell-1}^{m'} + \mathbf{e}_y \sum_{\substack{m'=\pm 1,\pm 3,\ldots \\ |m'|<\ell-1}} c_{\ell m;\ell-1,m'}^{y(1)} Y_{\ell-1}^{m'} + \mathbf{e}_z \sum_{\substack{m'=\pm 2,\pm 4,\ldots \\ |m'|<\ell-1}} c_{\ell m;\ell-1,m'}^{z(1)} Y_{\ell-1}^{m'}\Bigg] \\ +j_{\ell+1}(\rho)\Bigg[\mathbf{e}_x \sum_{\substack{m'=\pm 1,\pm 3,\ldots \\ |m'|<\ell+1}} c_{\ell m;\ell+1,m'}^{x(2)} Y_{\ell+1}^{m'} + \mathbf{e}_y \sum_{\substack{m'=\pm 1,\pm 3,\ldots \\ |m'|<\ell+1}} c_{\ell m;\ell+1,m'}^{y(2)} Y_{\ell+1}^{m'} + \mathbf{e}_z \sum_{\substack{m'=\pm 2,\pm 4,\ldots \\ |m'|<\ell+1}} c_{\ell m;\ell+1,m'}^{z(2)} Y_{\ell+1}^{m'}\Bigg], \text{ if } m \text{ is even} \\ j_{\ell-1}(\rho)\Bigg[\mathbf{e}_x \sum_{\substack{m'=\pm 2,\pm 4,\ldots \\ |m'|<\ell-1}} c_{\ell m;\ell-1,m'}^{x(1)} Y_{\ell-1}^{m'} + \mathbf{e}_y \sum_{\substack{m'=\pm 2,\pm 4,\ldots \\ |m'|<\ell-1}} c_{\ell m;\ell-1,m'}^{y(1)} Y_{\ell-1}^{m'} + \mathbf{e}_z \sum_{\substack{m'=\pm 1,\pm 3,\ldots \\ |m'|<\ell-1}} c_{\ell m;\ell-1,m'}^{z(1)} Y_{\ell-1}^{m'}\Bigg] \\ +j_{\ell+1}(\rho)\Bigg[\mathbf{e}_x \sum_{\substack{m'=\pm 2,\pm 4,\ldots \\ |m'|<\ell+1}} c_{\ell m;\ell+1,m'}^{x(2)} Y_{\ell+1}^{m'} + \mathbf{e}_y \sum_{\substack{m'=\pm 2,\pm 4,\ldots \\ |m'|<\ell+1}} c_{\ell m;\ell+1,m'}^{y(2)} Y_{\ell+1}^{m'} + \mathbf{e}_z \sum_{\substack{m'=\pm 1,\pm 3,\ldots \\ |m'|<\ell+1}} c_{\ell m;\ell+1,m'}^{z(2)} Y_{\ell+1}^{m'}\Bigg], \text{ if } m \text{ is odd.} \end{cases}$$

(A-24)

Thus, we have shown that the vector spherical harmonic functions ($\mathbf{M}_{o\ell m}, \mathbf{N}_{e\ell m}$) are linear combinations of the scalar spherical harmonic functions ($j_\ell Y_\ell^m$) in the Cartesian frame.

For the special case, when we consider the scattering of a plane wave from a sphere, only $m = 1$ component needs to be considered. Hence, Eq. (A-8), (A-11), (A-12) and Eq. (A-20)-(A-22) reduce to:

$$M_{o\ell 1}^{x} = \begin{cases} -\dfrac{1}{2}\ell(\ell+1)\sqrt{\dfrac{4\pi}{2\ell+1}}Y_{\ell}^{0}j_{\ell}(\rho), \ell=1 \\ \left[-\dfrac{1}{4}\sqrt{\dfrac{4\pi(\ell+2)(\ell+1)\ell(\ell-1)}{2\ell+1}}j_{\ell}(\rho)Y_{\ell}^{2} - \dfrac{1}{4}\sqrt{\dfrac{4\pi(\ell+2)(\ell+1)\ell(\ell-1)}{2\ell+1}}j_{\ell}(\rho)Y_{\ell}^{-2}, \right. \\ \left. -\dfrac{1}{2}\ell(\ell+1)\sqrt{\dfrac{4\pi}{2\ell+1}}j_{\ell}(\rho)Y_{\ell}^{0}\right], \ell \geq 2 \end{cases}$$

(A-25)

$$M_{o\ell 1}^{y} = \begin{cases} 0, \ell=1 \\ -\dfrac{1}{4i}\sqrt{\dfrac{4\pi(\ell+2)(\ell+1)\ell(\ell-1)}{2\ell+1}}j_{\ell}(\rho)Y_{\ell}^{2} + \dfrac{1}{4i}\sqrt{\dfrac{4\pi(\ell+2)(\ell+1)\ell(\ell-1)}{2\ell+1}}j_{\ell}(\rho)Y_{\ell}^{-2}, \ell \geq 2 \end{cases}$$

,

(A-26)

$$M_{o\ell 1}^{z} = -\dfrac{1}{2}\sqrt{\dfrac{4\pi\ell(\ell+1)}{2\ell+1}}Y_{\ell}^{1}j_{\ell}(\rho) + \dfrac{1}{2}\sqrt{\dfrac{4\pi\ell(\ell+1)}{2\ell+1}}Y_{\ell}^{-1}j_{\ell}(\rho), \ell \geq 1.$$

(A-27)

and

$$N_{e\ell 1}^{x} = c_{\ell-1,0}^{x(1)}j_{\ell-1}(\rho)Y_{\ell-1}^{0} + c_{\ell+1,0}^{x(2)}j_{\ell+1}(\rho)Y_{\ell+1}^{0} + c_{\ell+1,2}^{x(2)}j_{\ell+1}(\rho)Y_{\ell+1}^{2} + c_{\ell+1,-2}^{x(2)}j_{\ell+1}(\rho)Y_{\ell+1}^{-2}, \ell=1,2$$

$$c_{\ell-1,0}^{x(1)} = -\dfrac{\ell(\ell+1)^{2}}{2\ell+1}\sqrt{\dfrac{\pi}{2\ell-1}}$$

$$c_{\ell+1,0}^{x(2)} = \dfrac{\ell^{2}(\ell+1)}{2\ell+1}\sqrt{\dfrac{\pi}{2\ell+3}}$$

$$c_{\ell+1,2}^{x(2)} = -\dfrac{\ell}{2\ell+1}\sqrt{\dfrac{\pi}{2\ell+3}}\sqrt{\left[2+\dfrac{(\ell-1)(\ell+4)}{2}\right]\left[3+\dfrac{(\ell-1)(\ell+4)}{2}\right]}$$

$$c_{\ell+1,-2}^{x(2)} = -\dfrac{\ell}{2\ell+1}\sqrt{\dfrac{\pi}{2\ell+3}}\sqrt{\left[2+\dfrac{(\ell-1)(\ell+4)}{2}\right]\left[3+\dfrac{(\ell-1)(\ell+4)}{2}\right]}$$

,

(A-28)

$$N_{e\ell 1}^{x} = c_{\ell-1,0}^{x(1)}j_{\ell-1}(\rho)Y_{\ell-1}^{0} + c_{\ell-1,2}^{x(1)}j_{\ell-1}(\rho)Y_{\ell-1}^{2} + c_{\ell-1,-2}^{x(1)}j_{\ell-1}(\rho)Y_{\ell-1}^{-2}$$
$$+ c_{\ell+1,0}^{x(2)}j_{\ell+1}(\rho)Y_{\ell+1}^{0} + c_{\ell+1,2}^{x(2)}j_{\ell+1}(\rho)Y_{\ell+1}^{2} + c_{\ell+1,-2}^{x(2)}j_{\ell+1}(\rho)Y_{\ell+1}^{-2}, \ell \geq 3$$

$$c_{\ell-1,2}^{x(1)} = \dfrac{\ell+1}{2\ell+1}\sqrt{\dfrac{\pi}{2\ell-1}}\sqrt{\left[2+\dfrac{(\ell-3)(\ell+2)}{2}\right]\left[3+\dfrac{(\ell-3)(\ell+2)}{2}\right]}$$ ,

(A-29)

$$c_{\ell-1,-2}^{x(1)} = \dfrac{\ell+1}{2\ell+1}\sqrt{\dfrac{\pi}{2\ell-1}}\sqrt{\left[2+\dfrac{(\ell-3)(\ell+2)}{2}\right]\left[3+\dfrac{(\ell-3)(\ell+2)}{2}\right]}$$

$$N^y_{e\ell 1} = c^{y(2)}_{\ell+1,2} j_{\ell+1}(\rho) Y^2_{\ell+1} + c^{y(2)}_{\ell+1,-2} j_{\ell+1}(\rho) Y^{-2}_{\ell+1}, \ell = 1, 2$$

$$c^{y(2)}_{\ell+1,2} = \frac{i\ell}{2\ell+1}\sqrt{\frac{\pi}{2\ell+3}}\sqrt{\left[2+\frac{(\ell-1)(\ell+4)}{2}\right]\left[3+\frac{(\ell-1)(\ell+4)}{2}\right]} \quad , \qquad \text{(A-30)}$$

$$c^{y(2)}_{\ell+1,-2} = -\frac{i\ell}{2\ell+1}\sqrt{\frac{\pi}{2\ell+3}}\sqrt{\left[2+\frac{(\ell-1)(\ell+4)}{2}\right]\left[3+\frac{(\ell-1)(\ell+4)}{2}\right]}$$

$$N^y_{e\ell 1} = c^{y(1)}_{\ell-1,-2} j_{\ell-1}(\rho) Y^{-2}_{\ell-1} + c^{y(1)}_{\ell-1,2} j_{\ell-1}(\rho) Y^2_{\ell-1} + c^{y(2)}_{\ell+1,-2} j_{\ell+1}(\rho) Y^{-2}_{\ell+1} + c^{y(2)}_{\ell+1,2} j_{\ell+1}(\rho) Y^2_{\ell+1}, \ell \geq 3$$

$$c^{y(1)}_{\ell-1,2} = -\frac{i(\ell+1)}{2\ell+1}\sqrt{\frac{\pi}{2\ell-1}}\sqrt{\left[2+\frac{(\ell-3)(\ell+2)}{2}\right]\left[3+\frac{(\ell-3)(\ell+2)}{2}\right]} \quad ,$$

$$c^{y(1)}_{\ell-1,-2} = \frac{i(\ell+1)}{2\ell+1}\sqrt{\frac{\pi}{2\ell-1}}\sqrt{\left[2+\frac{(\ell-3)(\ell+2)}{2}\right]\left[3+\frac{(\ell-3)(\ell+2)}{2}\right]}$$

$$\text{(A-31)}$$

$$N^z_{e\ell 1} = c^{z(1)}_{\ell-1,1} j_{\ell-1}(\rho) Y^1_{\ell-1} + c^{z(1)}_{\ell-1,-1} j_{\ell-1}(\rho) Y^{-1}_{\ell-1} + c^{z(2)}_{\ell+1,1} j_{\ell+1}(\rho) Y^1_{\ell+1} + c^{z(2)}_{\ell+1,-1} j_{\ell+1}(\rho) Y^{-1}_{\ell+1}$$

$$c^{z(1)}_{\ell-1,1} = \frac{(\ell+1)^2 \sqrt{\ell(\ell-1)}}{2\ell+1}\sqrt{\frac{\pi}{2\ell-1}}$$

$$c^{z(1)}_{\ell-1,-1} = -\frac{(\ell+1)^2 \sqrt{\ell(\ell-1)}}{2\ell+1}\sqrt{\frac{\pi}{2\ell-1}} \qquad . \qquad \text{(A-32)}$$

$$c^{z(2)}_{\ell+1,1} = \frac{\ell^2 \sqrt{(\ell+1)(\ell+2)}}{2\ell+1}\sqrt{\frac{\pi}{2\ell+3}}$$

$$c^{z(2)}_{\ell+1,-1} = -\frac{\ell^2 \sqrt{(\ell+1)(\ell+2)}}{2\ell+1}\sqrt{\frac{\pi}{2\ell+3}}$$

## Table and Figure caption

**Table:**

[1] Best-fit parameters used in the theoretical modeling for Au nanoparticles without and with clusters.

**Figure:**

[1] Schematic drawing of (a) a chain of nanoparticles (top view) and close-packed clustering nanoparticles such as (b) trimer and (c) heptamers (top view).

[2] The electric field (at the origin as indicated in Fig. 1 with $z=0$), $|\mathbf{E}|$ as a function of photon energy for light scattering from a trimer of Au nanospheres of the same diameter $d=80nm$ (Fig. 1(b) and set $\varphi_d=0$) with two gaps: (a) large gap $g=10nm$ and (b) small gap $g=2nm$ calculated by the present Green's function method (dashed curves) with (a) ($\ell_{max}$, $N_k$, $N_z$) = (6,101,100) and (b) ($\ell_{max}$, $N_k$, $N_z$) = (9,121,100) and the extended Mie scattering theory (solid curves).

[3] The calculated reflectance for s- and p- polarized light for isolated clusters of Au-NPs (with $d=80$nm and set $\varphi_d=0$ in Fig. 1) on glass substrate with three different angles of incidence: 55° (green), 60° (red) and 65° (blue) for specular reflection for five different arrangements: (a) a chain of two nanoparticles (2) a chain of three nanoparticles (3) a chain of four nanoparticles (4) a trimer and (5) a heptamer. The gap used is 2nm.

[4] The calculated reflectance for s- and p- polarized light for isolated clusters of Au-NPs (with $d=80$nm and $\varphi_d=0$) on glass substrate with three different angles of incidence: 55° (green), 60° (red) and 65° (blue) for off-specular reflection for five different arrangements: (a) a chain of two nanoparticles (2) a chain of three nanoparticles (3) a chain of four nanoparticles (4) a trimer and (5) a heptamer. The gap used is 2nm.

[5] The orientation-averaged ellipsometric parameters, $\Psi$ and $\Delta$ as functions of photon energy obtained by the Green's function method for a random distribution of clustering nanoparticles for five different arrangements: (a) chains of two nanoparticles (2) chains of three nanoparticles (3) chains of four nanoparticles (4) trimmers, and (5) heptamers for three different angles of incidence: 55° (solid line), 60° (dashed line) and 65° (dash-dotted line) on the substrate. The particle size is 80nm and the gap is 2nm.

[6] Spectroscopic ellipsometry measurement (solid curves) and model calculations (dash-dotted curves) of randomly distributed Au nanoparticles with various arragmements. The nominal sizes of nanoparticles are (a) 20, (b) 40, (c) 60, and (d) 80 nm for incident angles of 55°, 60°, and 65°.

[7] Spectroscopic ellipsometry measurement (solid curves) and model calculations (dash-dotted curves) of random distribution of Au nanoparticles including the effect of clusters which are modeled by aggregations of nanoparticles with various arrangements. The nominal sizes of nanoparticles are (a) 40, (b) 60, and (c) 80 nm for incident angles of 55°, 60°, and 65°.

| Nanoparticle diameter $D$ (nm) | Similarity factor $f$ | Average pitch $p$ (nm) | Fraction of small clusters, $f_c$ | Fraction of nanoparticles in patches, $f_p$ | MSE for $\Psi$ ($\Delta$) noncluster model | MSE for $\Psi$ ($\Delta$) cluster model |
|---|---|---|---|---|---|---|
| 20 | 1.0 | 50 |  |  | 0.87 (9.89) |  |
| 36 | 0.8 | 140 | 0.015 | 0.005 | 1.27 (11.44) | 0.90 (9.68) |
| 60 | 0.7 | 170 | 0.02 | 0.05 | 1.81 (12.36) | 1.12 (9.87) |
| 80 | 0.7 | 245 | 0.025 | 0.05 | 2.38 (13.84) | 1.53 (10.79) |

**Table. 1**

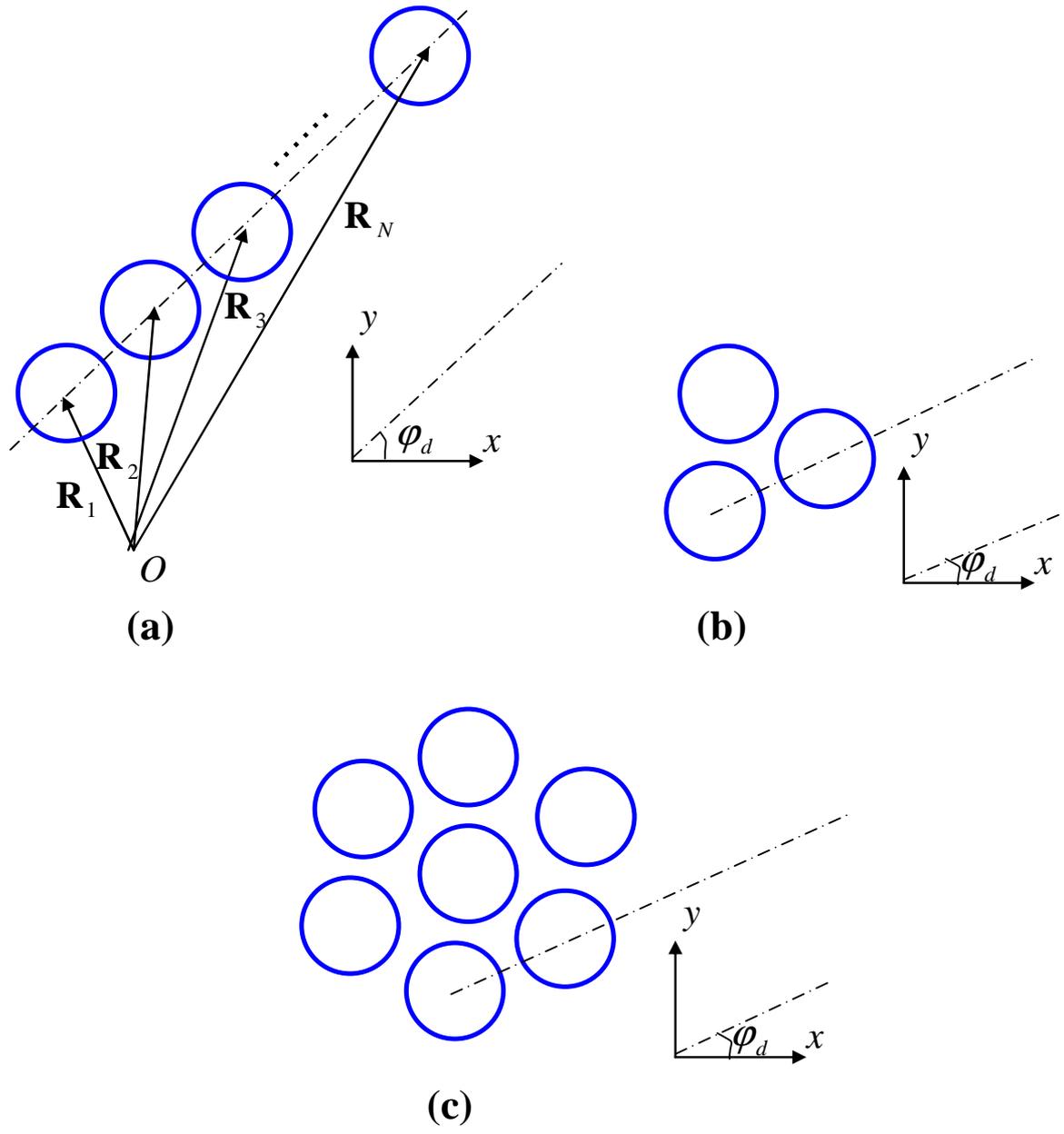

**Fig. 1**

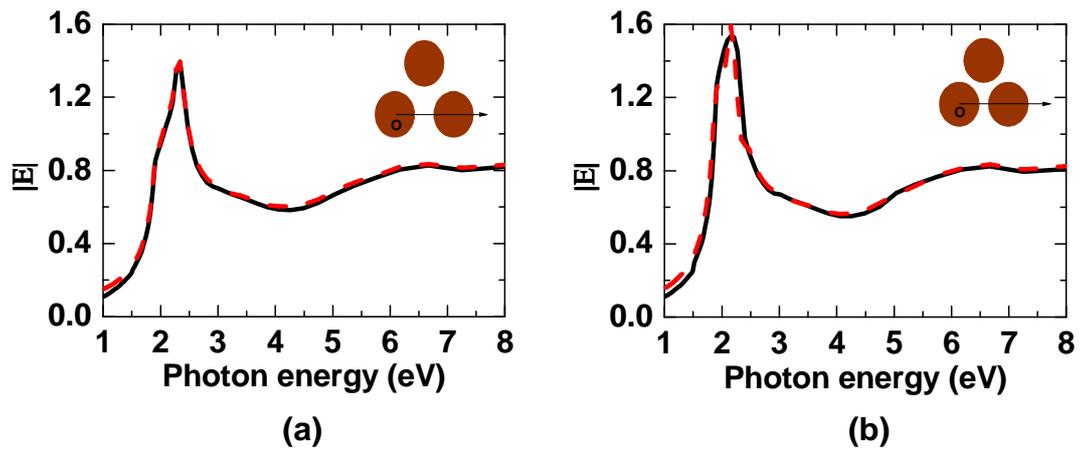

**Fig. 2**

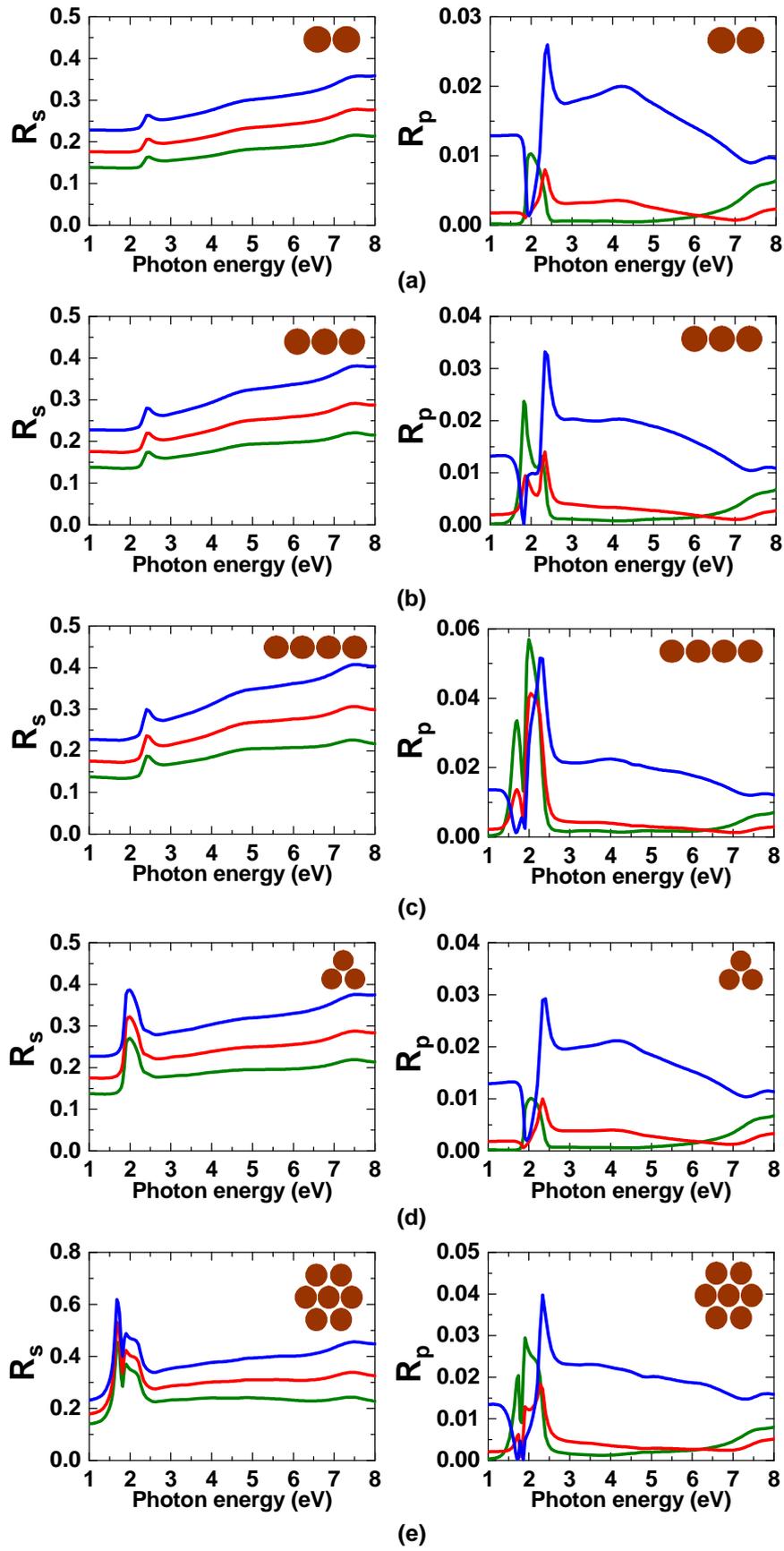

Fig. 3

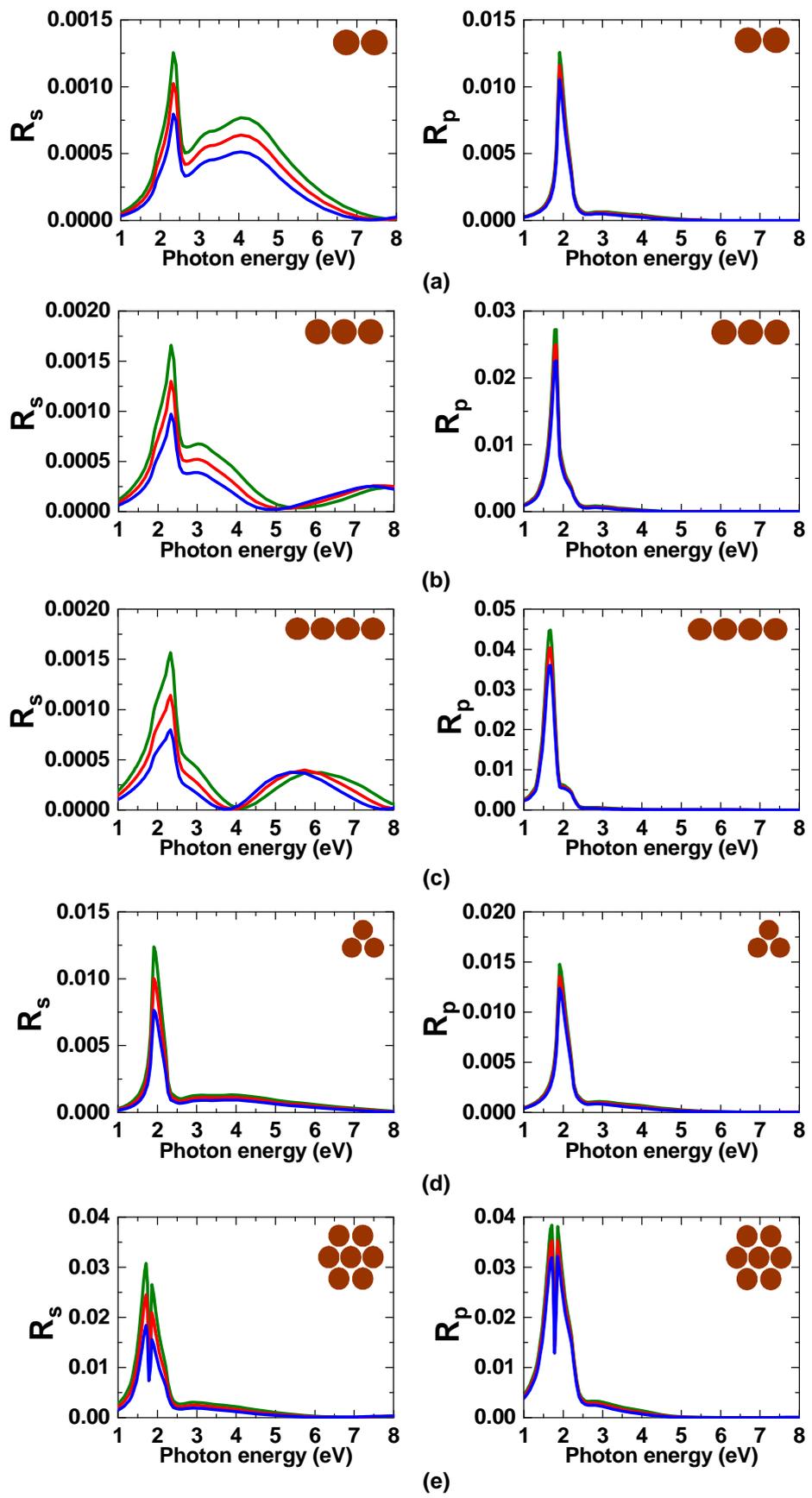

Fig. 4

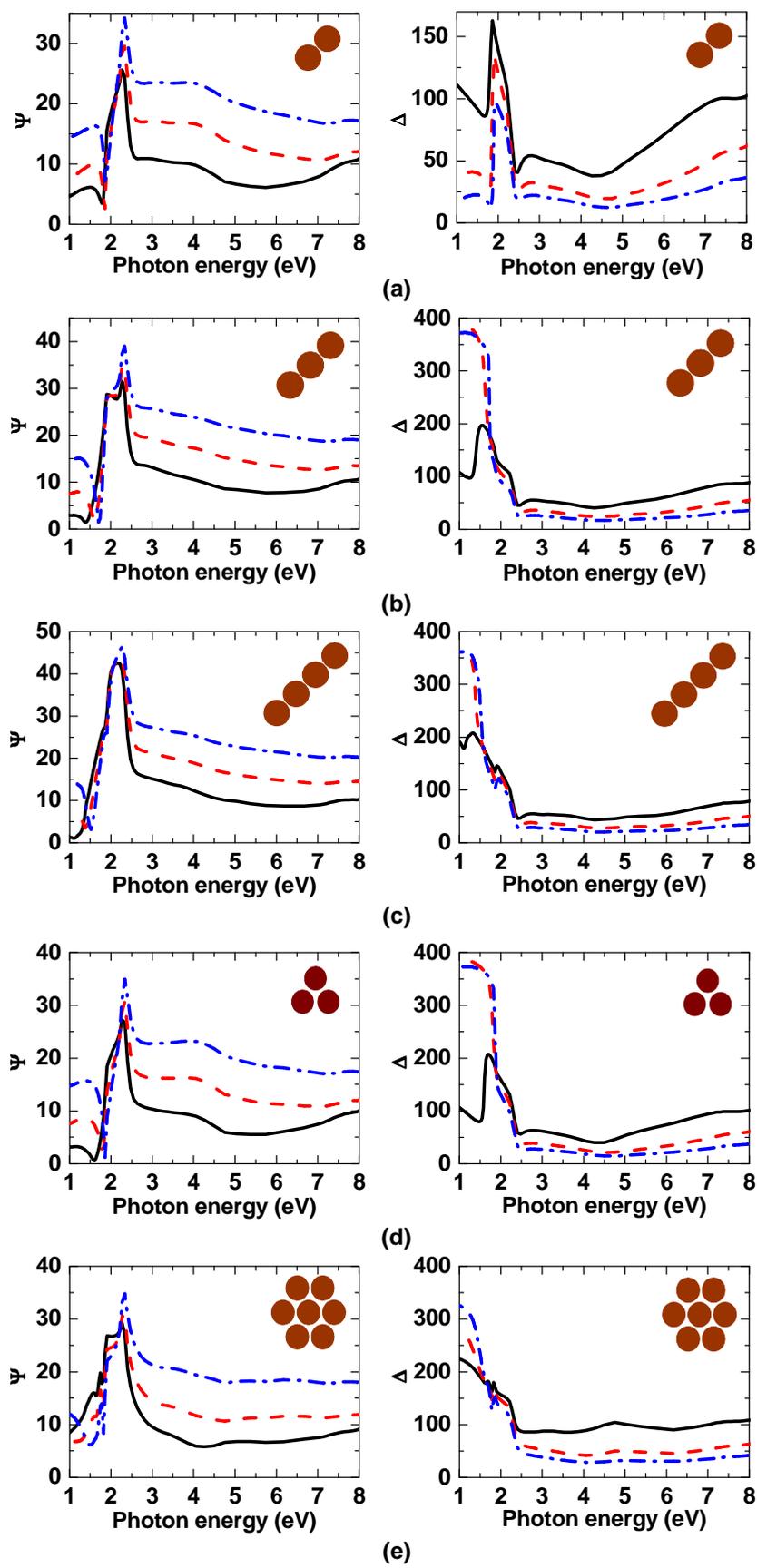

Fig. 5

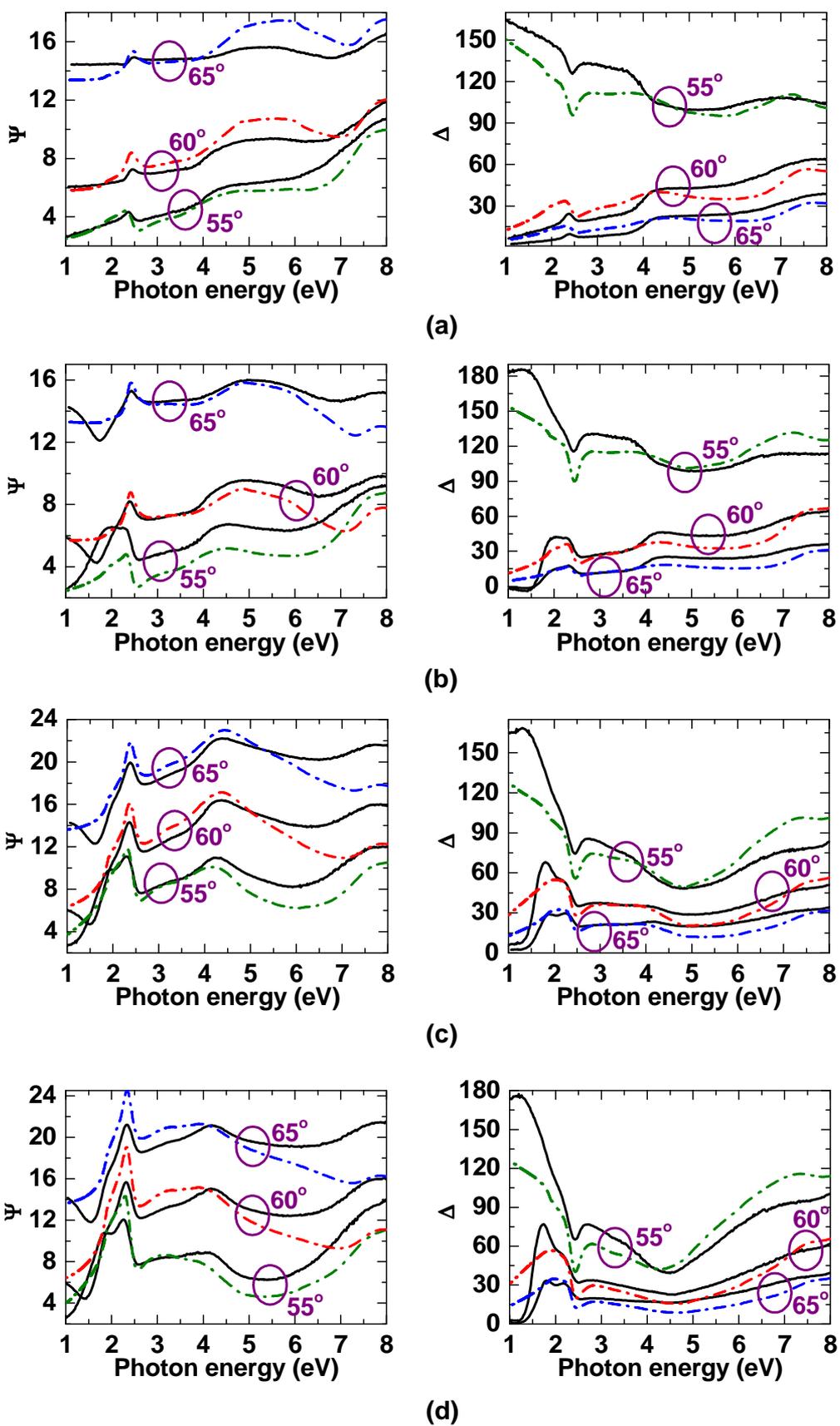

**Fig. 6**

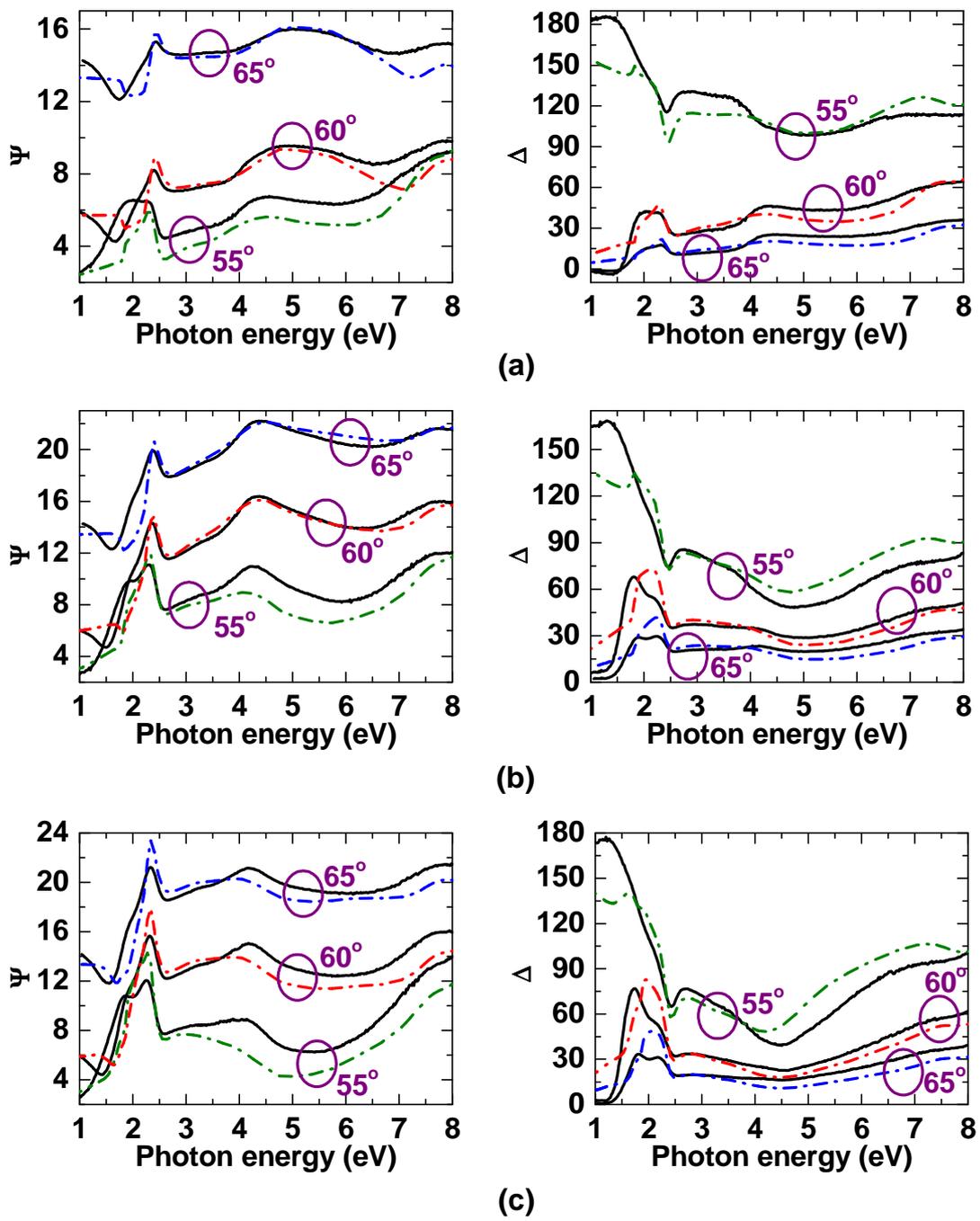

Fig. 7